\documentclass{article}

\usepackage{microtype}
\usepackage{graphicx}
\usepackage{subfigure}
\usepackage{booktabs} % for professional tables

\usepackage[hyphens]{url}
\usepackage{hyperref}

\usepackage[accepted]{icml2025}

\usepackage{amsmath}
\usepackage{amssymb}
\usepackage{mathtools}
\usepackage{amsthm}

\usepackage[capitalize,noabbrev]{cleveref}

\theoremstyle{plain}

\theoremstyle{definition}

\theoremstyle{remark}

\usepackage[textsize=tiny]{todonotes}

\icmltitlerunning{Position: Ensuring mutual privacy is necessary for effective external evaluation of proprietary AI systems}

\begin{document}

\twocolumn[
\icmltitle{Position: Ensuring mutual privacy is necessary for effective external evaluation of proprietary AI systems}



\icmlsetsymbol{equal}{*}

\begin{icmlauthorlist}
\icmlauthor{Ben Bucknall}{eng,aigi}
\icmlauthor{Robert F. Trager}{aigi,bsg}
\icmlauthor{Michael A. Osborne}{eng,aigi}
\end{icmlauthorlist}

\icmlaffiliation{eng}{Department of Engineering Science, University of Oxford}
\icmlaffiliation{bsg}{Blavatnik School of Government, University of Oxford}
\icmlaffiliation{aigi}{Oxford Martin AI Governance Initiative}

\icmlcorrespondingauthor{Ben Bucknall}{bucknall@robots.ox.ac.uk}

\icmlkeywords{Machine Learning, ICML}

\vskip 0.3in
]



\printAffiliationsAndNotice{}  % leave blank if no need to mention equal contribution

\begin{abstract}
The external evaluation of AI systems is increasingly recognised as a crucial approach for understanding their potential risks.
However, facilitating external evaluation in practice faces significant challenges in balancing evaluators' need for system access with AI developers' privacy and security concerns.
Additionally, evaluators have reason to protect their own privacy – for example, in order to maintain the integrity of held-out test sets.
We refer to the challenge of ensuring both developers’ and evaluators’ privacy as one of providing \emph{mutual privacy}.
In this position paper, we argue that
(i) addressing this mutual privacy challenge is essential for effective external evaluation of AI systems, and
(ii) current methods for facilitating external evaluation inadequately address this challenge, particularly when it comes to preserving evaluators’ privacy.
In making these arguments, we formalise the mutual privacy problem; examine the privacy and access requirements of both model owners and evaluators; and explore potential solutions to this challenge, including through the application of cryptographic and hardware-based approaches.

\end{abstract}

\section{Introduction}
External evaluations -- including independent audits and red-teaming -- are emerging as a cornerstone of AI governance regimes.
Their importance is reflected in their inclusion as Measure 16 of the second draft of the AI Act's \emph{General-Purpose AI Code of Practice} \cite{oliver_second_2024}, and prior research has demonstrated their value in advancing external transparency and scrutiny of advanced AI systems \cite{raji_outsider_2022, mokander_auditing_2023, anderljung_towards_2023}.
For external evaluations to be effective, evaluators require sufficient access to both the AI system being assessed and relevant contextual information regarding that system \cite{bucknall_structured_2023, casper_black-box_2024}.
However, in the case of proprietary systems -- where model weights are not made openly available -- this can create a practical tension between providing external access and maintaining system security, leading many model providers to be hesitant about granting deep access to evaluators.

Current research on expanding system access for external evaluators has primarily focused on balancing evaluator access with model providers' security concerns.
However, equally crucial but often overlooked is the privacy and security of the evaluators themselves.
When evaluators lack adequate privacy protections, it can be difficult to fully trust either the independence or results of their evaluations.

As such, in this paper we argue for the following pair of positions:
\begin{itemize}
    \item \textbf{Ensuring the mutual privacy of both external auditors and model owners is essential for effective external evaluation of AI systems; and}
    \item \textbf{Current methods for facilitating external evaluations of proprietary systems inadequately address the mutual privacy challenge, particularly when it comes to preserving evaluators' privacy.}
\end{itemize}

We make three key contributions in advocating for these positions.
First, we introduce and define the challenge of ensuring mutual privacy during external evaluations of AI systems. 
Second, we provide an overview of information that both model owners and evaluators may need to keep private or share during evaluations.
Third, we discuss current and proposed methods for enabling secure external evaluations, finding that evaluator privacy is rarely considered.

\section{Formalising the mutual privacy problem}
We start by formalising the notion of a \textit{mutual privacy problem} using set notation. This formalism is given solely for increasing the precision and clarity of the definition.
% We do not use set theory to as a language to formally prove properties of the system, though this could be an interesting avenue for future work.

Suppose that we have two actors, a model owner $M$, and an external evaluator $E$. Each actor has a set of information known to them, denoted $I_M$ and $I_E$, respectively, where the intersection $I_M \cap I_E$ may be nonempty -- that is, there may be common knowledge.
For each actor $i \in \{M, E\}$, there is some subset $P_i \subset I_i$, which we call $i$'s private set, that they wish to keep private from the other -- that is, $P_M \cap I_E = \emptyset$, and vice-versa.
Furthermore, each actor has a set of \emph{required information}, $R_i \subset I_j$ containing information that may be demanded from the other for the purposes of conducting or facilitating an external evaluation.
We say that there is a \emph{privacy problem} if $P_M \cap R_E \neq \emptyset$ -- that is, if the model owner wishes to know something that the evaluator would like to keep private -- or vice-versa. A \textit{mutual} privacy problem occurs if both $P_M \cap R_E$ and $P_E \cap R_M$ are nonempty.

We take a fairly inclusive conceptualisation for what constitutes `information', including not only factual knowledge (for example, architectural details of a given AI model), but also procedural knowledge (for example, information about what activities are being carried out by the other actor, and how), as well as access to system artefacts including models and datasets.
Information could also refer to assurances, for example that the model owner's private set $P_M$ is indeed private and not known to the evaluator, or that the information $R_E$ that the evaluator receives from the model owner is accurate and reliable.

\section{Information that is private to, and required by, each actor}

\begin{figure*}
    \centering
    \includegraphics[width=\linewidth]{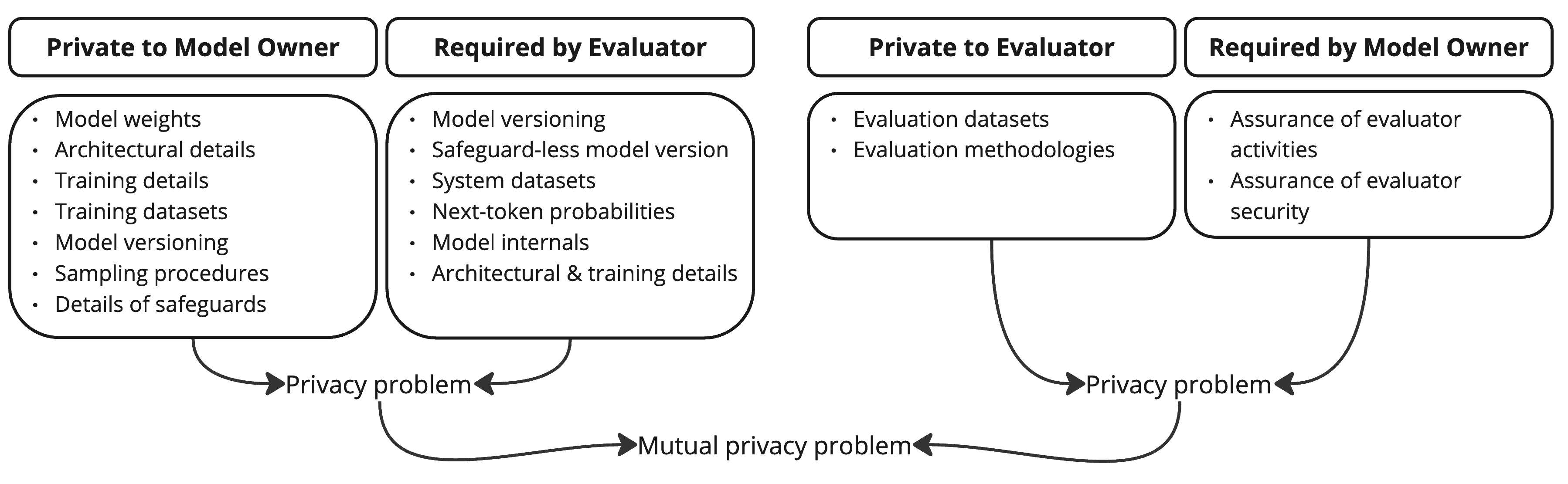}
    \caption{A summary of the relevant information that either the model owner or evaluator may require or want to keep private. Where one party requires information that the other would wish to keep private, we say there is a `privacy problem'. Where both parties require information that the other would wish to keep private, we say there is a `\emph{mutual} privacy problem.'}
    \label{fig:summary}
\end{figure*}

In this section we analyse a typical external evaluation setting in terms of the information that each of the model owner and evaluator both require and want to keep private. For each set of information we give a brief description, along with an explanation of why this information is may need to be private and/or required.
Note that we do not claim that all of the below will hold in \emph{all} instances of external evaluation. The extent and frequency with which the types of information described below will be required and/or private varies. Rather, we refer to information for which, in some plausible cases, there may be \emph{some} reason to include in any of $P_M$, $P_E$, $R_M$, or $R_E$ for the purpose of conducting an external evaluation. %for the model owner or external evaluator to keep this information private, or request access to it.
A summary of the information discussed and how conflicting requirements can give rise to a mutual privacy problem is depicted in Figure \ref{fig:summary}.

\subsection{Information private to the model owner}
\label{sec:owner_private}
We begin by describing information known to the model owner that they may want to keep private, either from the external evaluator, or the broader public. For each set of information we provide a brief explanation of the potential incentives for this privacy.

\textbf{Model weights.}
The most advanced contemporary commercial AI models cost on the order of tens of millions of dollars to train, placing considerable value on the resulting model weights \cite{cottier_rising_2024}. 
Therefore, unless pursuing an open-weights commercial strategy \cite{widder_why_2024}, model owners have large incentives to ensure that model weights are not leaked or stolen, motivating strict controls on who can access the raw weight data.
Furthermore, we are seeing an increasing number of statements regarding the importance of ensuring the security of model weights for reasons of national security and as a means of managing misuse risks \cite{nevo_securing_2024, openai_reimagining_2024, us_ai_safety_institute_managing_2024}.
This motivates the need to keep model weights secure and private to the model owner, and even potentially a subset of individuals within the organisation.
Furthermore, to the extent that other model internals, such as activations, can be used to extract or reconstruct the model, there is incentive to keep such internals private.

\textbf{Architectural details.}
Architectural details refer to the `internal structure' of the model, either at the macro- (for example, whether the model is based on a transformer architecture \cite{vaswani_attention_2017}, a diffusion model \cite{sohl-dickstein_deep_2015}, etc.) or micro level (for example, the number, dimensions, and types of layers used -- e.g. MLP, attention \cite{bahdanau_neural_2014}, mixture of experts \cite{jacobs_adaptive_1991, eigen_learning_2013}, etc.).
Architectural innovations contribute to overall algorithmic improvements, and have been estimated to half the amount of compute required to meet a set performance threshold approximately every eight months \cite{ho_algorithmic_2024}.\footnote{See \cite{erdil_how_2025} for a discussion of the improvements made by DeepSeek-v3 \cite{deepseek-ai_deepseek-v3_2024} to the Transformer architecture -- specifically, \emph{multi-head latent attention} (MLA) and \emph{DeepSeekMoE}.}
This has the parallel effect of increasing the performance able to be achieved from a given quantity of computational resources \cite{pilz_increased_2023}.
As such, model owners frequently omit architectural details from technical reports of new models due to \textit{`the competitive landscape'} \cite{openai_gpt-4_2023}.\footnote{Compare the lack of architectural details in the GPT-4 technical report \cite{openai_gpt-4_2023} to those given in the GPT-3 paper \cite{brown_language_2020}.}
In addition, concerns regarding the privacy of model architecture could incentivise restricting other forms of access that could be used to infer architectural details, including, for example, the ability to view output logits or probabilities \cite{carlini_stealing_2024}.

\textbf{Training details.}
Training details refer to the algorithmic and engineering methods used during the pretraining and fine-tuning stages of a model's development.
As in the case of architectural details discussed above, advances in training methodologies are responsible for attaining greater system performance without scaling the compute or datasets used in training \cite{ho_algorithmic_2024}.
For example, FlashAttention \cite{dao_flashattention_2022} introduced optimisations for computing attention, leading to a claimed 7.6x speedup.
As such, model owners may have incentives to not disclose advances in training methodologies so as not to provide resulting benefits to commercial competitors.

\textbf{Training datasets.}
There are many reasons for why model owners may have incentives to keep the training datasets used in the development of their model(s) private.
Firstly, training datasets can be costly to assemble and collate, especially those based on human feedback, for example for training a reward model when performing reinforcement learning from human feedback (RLHF) \citep[See e.g.,][sec. 3.1.3]{casper_open_2023}, providing a commercial incentive to keep these datasets private.
Furthermore, datasets may contain private, sensitive, or inappropriate material \cite{luccioni_whats_2021, thiel_identifying_2023, birhane_into_2023, birhane_multimodal_2021} which could cause harm if released publicly, or otherwise leaked.

In some cases, pretraining datasets may contain large amounts of user data from other services provided by a model owner (or a parent company) -- for example, Meta having access to Facebook and Instagram users' content or Google DeepMind having access to YouTube data for the purposes of training.
Putting aside the legitimate ethical concerns surrounding the use of such data for training AI systems \citep[see e.g.,][]{noyb_noyb_2024}, model owners that have access to non-public data sources may be unwilling to share training datasets derived from them due to the competitive advantage that doing so would forfeit.
Furthermore, even in the absence of commercial incentives, the public release or private transfer of such data may face legal opposition in light of data handling regulations such as the EU's General Data Protection Regulation (GDPR).
Finally, model owners may have an incentive to keep datasets private in order to use opacity to defend themselves from copyright liability \cite{henderson_foundation_2023}, or data handling malpractice.

\textbf{Model versioning.}
A model owner may have commercial incentives to obscure details regarding the precise version of a model that is being evaluated or deployed to users.
This may be due to an inherent tension between the commercial utility of a system, and its performance in external evaluations or audits, creating incentives for a model owner to `swap out' model versions between each situation in order to attain better apparent performance in each.
Additionally, being able to precisely specify a model version could inadvertently disclose other types of information discussed here, such as architecture or details of post-training enhancements.

Relatedly, commercial models almost always undergo some amount of safety fine-tuning (or other modifications) before release, as well as other modifications that make the model more helpful to users.
However, non safety-tuned model versions do exist, even if not commercially released.
Given that these models do not have safeguards (e.g. to refuse harmful or toxic requests), model owners will have an incentive to limit access to them in order to reduce potential downstream harm and related reputational risks.

\textbf{Sampling procedures.}
Recent high-profile releases have signaled a move to `scaling up' inference compute through more complex sampling procedures as a way of increasing model performance at deployment time \cite{brown_large_2024, davidson_ai_2023, villalobos_trading_2023, lee_evolving_2025, patel_scaling_2024}.\footnote{While details regarding the precise methods used in OpenAI's recent o1 \cite{openai_openai_2024} and o3 models are closely guarded, see some speculative discussion in \cite{chollet_openai_2024}.}
Thus, as in the case of details pertaining to both the architecture and training methods of commercial models, developers will have an incentive to keep information regarding sampling procedures out of public view so as not to diminish any competitive advantage that advancements may confer.

\textbf{Details of implemented safety measures.}
Finally, model owners may also have incentives not to disclose information detailing any safety guardrails applied to a given model.
If leaked, information pertaining to the implemented guardrails may could be used to more effectively jailbreak or otherwise bypass the safeguards, increasing the  risk of downstream harm stemming from misuse, as well as creating reputational risk to the model owner.
Furthermore, since in many cases, more robust safeguards can increase the `deployability' a model, information regarding employed safeguards may constitute sensitive IP that could advantage commercial competitors if publicly leaked.

\subsection{Information required by the evaluator}
\label{sec:evaluator_required}
Here we describe and motivate information and access that external evaluators may require in order to effectively carry out an evaluation of a proprietary system.

\textbf{Model versioning.}
When interpreting the results of external evaluations, it is informative to have clarity regarding the precise model version on which the evaluation was conducted -- for example, to be able to compare two evaluations conducted on the same model, and whether the evaluated version is the same as that which is made commercially available to end-users. % Also important for maintaining reproducibility and replicability? though this also depends on the same version being available -- just knowing that it's changed doesn't really help with this
In the absence of credible model version identification, model owners could potentially `swap out' model versions between pre-deployment tests and commercial deployment, such that results of the evaluation do not generalise to the version available to end-users.
While this could potentially involve the disclosure of system details, this may not be necessary in all cases. For example, a system could be implemented by which each model version has an arbitrary ID that does not convey any information about that model, but nonetheless serves as a unique version identifier.
Finally, while being able to precisely identify individual model versions is valuable for external evaluators, it may also be desirable to more publicly disclose identifiers so as to foster trust throughout the ecosystem.

\textbf{Access to a model version without safeguards.}
Many evaluations aim to assess the maximal, upper-bound capabilities of AI systems, potentially on harmful or dual-use tasks \citep[see e.g.,][sec. 5]{uk_ai_safety_institute_early_2024}.
In order to get an accurate representation of these upper-bound capabilities it is useful for evaluators to have access to a `helpful-only' version of the model that has not undergone post-training modifications aimed at reducing the model's propensity to output harmful content \citep[][sec. 7]{uk_ai_safety_institute_early_2024}.
% Furthermore, having access to a model that does not have post-training safeguards applied allows for evaluation of the efficacy of such safeguards by means of comparison with a version that does include safeguards.
Furthermore, access to versions of a model both with and without post-training safeguards enables direct evaluation of those safeguards' effectiveness through comparative approaches.

\textbf{System datasets.}
Datasets required for the purpose of external evaluation could include those used in pretraining, fine-tuning, or retrieval.
Datasets are frequently the subject of audits and evaluation in their own right \citep[see e.g.,][]{thiel_identifying_2023, birhane_into_2023, birhane_multimodal_2021}. In such cases, some form of access to, and information about, the dataset is clearly necessary.
Though this may not necessarily require `full', unrestricted access to inspect the dataset at the level of individual datapoints, the level of thoroughness of the evaluation is likely impacted by the level of access and information granted.
For example, while being able to query a dataset to ascertain whether it contains a given datapoint could be useful for identifying copyright violations, it is likely insufficient for more exploratory analysis aiming to comprehensively identify harmful content.

Even in cases where datasets are not the subject of an evaluation, they can provide instrumental value by providing greater context regarding results of model evaluations.
For example, being able to distinguish whether a capability exhibited by a model during testing is an instance of generalised reasoning, or extrapolation from similar training examples, depends on being able to inspect the contents of a model's training dataset. As an anecdote, GPT-4's ability to draw a unicorn in \emph{TikZ},\footnote{\emph{TikZ} is a package for drawing graphical elements in LaTeX.} as showcased in \cite{bubeck_sparks_2023}, appears impressive if you are unaware of the possibility of many such drawings being included in its training dataset.\footnote{See, for example, this thread on StackOverflow collating TikZ drawings of various animals: \url{https://tex.stackexchange.com/questions/387047/the-duck-pond-showcase-of-tikz-drawn-animals-ducks}.}
Additionally, being able to measure and the overlap between training and evaluation datasets is crucial for accurately interpreting evaluation results \cite{zhang_language_2024}.

\textbf{Logits or next-token probabilities.}
The ability to observe output logits or next-token probabilities produced by a model allows for an efficient reduction in the variance of evaluation results. This effectively bypasses the need to conduct resampling of the model to attain statistical mean performance on a given task, reducing the computational cost of conducting evaluations with statistical significance \citep[][sec. 3.2]{miller_adding_2024}.
Furthermore, next-token probabilities are necessary for calculating many standard performance metrics on benchmarks, such as the Brier score \cite{brier_verification_1950}, as well as being instrumental for predicting downstream performance based on scaling \cite{schaeffer_why_2024}.

\textbf{Model internals.}
Providing evaluators with white- and grey-box access to AI models, whereby some or all of a model's internals are able to be inspected, can enable the application of more powerful evaluation methodologies than if an evaluator is only able to observe a model's generated responses in a black-box setting \cite{casper_black-box_2024}.
Model internals could include model weights, intermediate activations, attention patterns, or gradients.
White- and grey-box access could also allow for the more efficient application of methods that are nonetheless able to be carried out in black-box settings, for example by automating a search for adversarial inputs \citep[][sec. 4.1]{zou_universal_2023, casper_black-box_2024}.

\textbf{Architectural \& training details.}
Evaluators can benefit from knowing certain details regarding the architecture of the system being interacted with, and how it was trained, as a means of informing hypotheses as to why certain behaviours or properties were observed.
This in turn can help inform the implementation of suitable mitigations, in cases where undesirable behaviours are found \cite{casper_black-box_2024}.
Furthermore, high-level knowledge, such as whether and how safety fine-tuning was carried out on a model, or whether the system performs hidden chain-of-thought reasoning, can enable evaluators to better calibrate judgements regarding the capabilities and vulnerabilities of that system.

\subsection{Information private to the evaluator}
\label{sec:evaluator_private}
Here we outline information known to the evaluator that may require being kept private from the model owner during an external evaluation.

\textbf{Evaluation datasets.}
For cases in which evaluators have standardised evaluation sets on which they assess the performance of systems, there are multiple reasons for wanting these sets to be kept private.
Here we raise two such reasons, namely i) in order to ensure minimal test-train overlap, and ii) to ensure the security of sensitive information. We discuss each in turn.

Many concerns have been raised regarding the importance of preserving a strict test-train split when evaluating models\footnote{Also referred to as preventing training data contamination.} -- something that is increasingly challenging in the case of language models that have been trained on large portions of the internet \cite{zhang_language_2024, jiang_investigating_2024, oren_proving_2023}.
Having samples from evaluation sets inadvertently included in training datasets drastically reduces the evaluation's utility for measuring a model's generalisation capabilities.
Thus, in order to maintain the validity and efficacy of their evaluations, evaluators will have a strong incentive to ensure that, in the course of evaluating a model, their evaluation sets are not being observed, logged, or otherwise recorded by the model owner, as this could open the door to the potential use of evaluation datasets during training.
This is particularly challenging when evaluating proprietary systems made available only through an application programming interface (API) operated by the model owner, and is motivation for OpenAI's o3 model only being evaluated on the \emph{`semi-private'} test set of the ARC-AGI benchmark, rather than the fully held-out set \cite{chollet_arc_2024, chollet_openai_2024}.

Secondly, in cases in which an evaluator is aiming to assess a model's performance in a domain with national-security relevance, for example cybersecurity, the evaluation set may contain sensitive and potentially harmful information \cite{thurnherr_who_2025, bricman_hashmarks_2023}.\footnote{See also, the UK AI Safety Institute's \emph{`partnering with government experts to directly assess the most national-security-relevant dangerous capabilities of models.'} \cite{uk_ai_safety_institute_advanced_2024}}
In these cases there will be further incentive to prevent leakage of the evaluation set, due to the potentially large-scale national security implications that could result from a leak.

\textbf{Evaluation methodologies and metrics.}
Even without direct access to observe samples contained in an evaluation dataset, there are reasons to keep the evaluation methodology or metrics private from the model owner so as to preserve the evaluation's integrity. 
As put by \citet{jones_under_2024}, \emph{`if evaluation metrics are made public, it incentivises developers to build their model to simply meet the metrics,'} potentially leading to case of \emph{Goodhart's law}.\footnote{\emph{Goodhart's law}, as commonly formulated, states that \emph{`when a measure becomes a target, it ceases to be a good measure.'}}

A potentially informative analogue here is that of vehicle emissions testing -- specifically, the case of the Volkswagen emissions scandal.\footnote{This comparison has been made elsewhere, for example in \cite{gal_towards_2024}.}
As described in the Environmental Protection Agency's (EPA) notice of violation letter, certain models of Volkswagen cars contained software that was able to \emph{`[sense] whether the vehicle [was] being tested or not based on various inputs including the position of the steering wheel, vehicle speed, the duration of the engine's operation, and barometric pressure'} which would then instruct the vehicle's \emph{`emission control systems to underperform when the software determined that the vehicle was not undergoing the federal test procedure.'}
Furthermore, \emph{[t]hese inputs precisely [tracked] the parameters of the federal test procedure used for emission testing for EPA certification purposes'} \cite{brooks_notice_2015}.
While there are doubtless reasons for openly publishing the federal emissions testing procedure, it is not unreasonable to hypothesise that, had Volkswagen not known the testing parameters in such detail, they would not have been able to implement software that could precisely predict whether the vehicle was undergoing testing.

In bringing this analogue back to the case of evaluating AI systems, there may be cases in which, if AI developers have visibility into the precise methods or metrics of an independent evaluation, they may be able to improve the resulting score in the narrow set of test cases, without improving performance on the underlying property that the evaluation is aiming to assess -- essentially artificially reducing the internal validity of the test.
As a simple demonstrative example, consider an assessment of the performance of a classification algorithm. If the model owner knows that an evaluator will be measuring a particular metric in a confusion matrix -- say, for example, true positive rate (TPR) -- then they could artificially boost TPR at the expense of other metrics.

Furthermore, while Volkswagen implemented a system whereby a vehicle automatically detected when it was being tested, the ways in which an AI developer could achieve the same outcome need not be as sophisticated.
In particular, they likely would not need any method of automatically detecting a test environment, as developers are usually aware of all tests taking place, and ultimately retain control of the system -- unlike in the case of  vehicles in others' possession.

\subsection{Information required by the model owner}
\label{sec:owner_required}
Here we outline information that would be required or desired by the model owner during external evaluations.

\textbf{Assurance that evaluator activities remain within predefined bounds.}
External evaluation, and particularly external red-teaming, involves the explicit definition of a legal safe harbour, within which evaluators are able to act without fear of reprimand for contravening the usual terms of use \cite{longpre_safe_2024}.
While safe harbours could be implemented so as to provide \emph{ex post} assurance that evaluators did not break the predetermined terms of agreement, the model owner may demand some amount of assurance that evaluators are not acting outside the bounds of this safe harbour while the evaluation is ongoing.
Providing this assurance may require transparency into the actions being performed by the evaluator.

\textbf{Assurance of evaluator security practices.}
Due to the strong incentives for ensuring the security of many types of information outlined in section \ref{sec:owner_private}, model owners may require assurance that external evaluators are upholding robust security practices as a way of minimising any leakage of the information that they may have privileged access to.
As in the case of confirming that evaluators are acting within predefined bounds, this may require providing model owners with significant levels of transparency into evaluator activities.

\section{Discussion}
Having laid out the preferences that both model owners and external evaluators may have for privacy and transparency, in this section we discuss where these preferences induce a mutual privacy problem. We further discuss why overcoming this problem is important for ensuring the efficacy of external evaluation, and why both current and proposed practices for conducting external evaluations are insufficient for ensuring mutual privacy.

\subsection{Exploring the mutual privacy challenge of external evaluation}
From the previous section we observe multiple conflicting incentives whereby a model owner has reason to keep information private that would be useful for an external evaluator assessing the model in question, and vice-versa. Firstly, we observe that details pertaining to the construction and development of the model, including regarding the architecture, training, and sampling procedures employed, are both commercially sensitive for the model owner and contextually useful for external evaluators.

Secondly, training datasets appear to be one of the more contentious points of conflict, being both sensitive to model owners for commercial, legal, and ethical reasons, as well as being a clearly necessary provision for an external dataset audit, and instrumentally valuable in the case of model evaluations.

Thirdly, model internals, such as weights and activations, represent some of the most sensitive information that model owners will want to keep private (again, assuming that they are not following an open-weight commercial strategy \cite{widder_why_2024}). While not strictly necessary for behavioural model evaluations, white- or grey-box access to internals can greatly improve the efficiency and robustness of external evaluations, and would be crucial for the application of interpretability-based evaluation techniques \cite{casper_black-box_2024}.

Finally, preserving the integrity of evaluation datasets employed by external evaluators poses a significant challenge in the other direction. While model owners may not have an explicit need to view the contents of such datasets, their incentives to ensure that external evaluators are acting within pre-agreed bounds and upholding sufficient security practices may indirectly motivate the direct observation of evaluator activities -- something that could reveal the contents of evaluation datasets.

Notably, the tensions between privacy and transparency appear in both directions -- with both model owners and evaluators requiring both privacy and transparency of a shared set of information. Thus, we observe that in many cases, there is indeed a problem of ensuring \emph{mutual} privacy.
In the absence of methods for assuring that both model owners' and evaluators' privacy and transparency requirements are met, the reliability of externally-conducted AI evaluations will be in question, either as a result of evaluators having insufficient information, or through a lack of assurance regarding the integrity of the evaluation itself.

It is worth also pointing out areas that seem less challenging and could be seen as first-steps that could be taken for improving the efficacy of external evaluations in the short term. For example, providing transparency into the precise model version that is being evaluated, and that it corresponds to that which is made available to users may be feasible without revealing specific details about the model. In the longer term, it may be possible to provide robust cryptographic guarantees that the model versions being evaluated and deployed correspond \citep[][sec. 5.4.1]{reuel_open_2024}.

Furthermore, providing evaluators with access to output logits or next-token probabilities can significantly aid in improving the statistical validity of conducted evaluations \cite{miller_adding_2024}. While, as observed above, this enables attacks that could extract undisclosed information about the model, such as the hidden dimension \cite{carlini_stealing_2024},
this risk could plausibly be addressed through comprehensive legal agreements between model owners and evaluators. More hypothetically, it may be possible to implement methods that can automatically detect extraction attacks based on the systematic sampling patterns required.

\subsection{Current practice in external evaluation is insufficient for ensuring mutual privacy}
Current external evaluation practices commonly involve third parties accessing proprietary systems through APIs operated and hosted by the model owner \citep[see e.g.,][app. D]{metr_details_2024, meinke_frontier_2024}.\footnote{Precise details regarding the access afforded to external evaluators are not disclosed, though see \cite{harrington_external_2024} for a high-level overview.}
While in some cases evaluators may be given certain privileges, for example through relaxed rate limits, subsidised API credits, or the ability to perform custom fine-tuning, such access is still largely similar to that given to users upon deployment, and thus does not afford access to much of the information required by evaluators as set out in section \ref{sec:evaluator_required}.

Evaluating models through a model owner-hosted API could in principle allow the owner to observe all evaluator activities, as a way of ensuring that evaluators are acting within pre-agreed bounds. However, this does not afford external evaluators any guarantee that their privacy concerns are being addressed beyond trusting model owners' claims that model interactions taking place through their API are not being logged.
While some developers do offer \emph{`zero data retention'} (ZDR) options to certain customers, it is not clear whether such options are employed in cases of external evaluation, nor whether it would be possible to verify that ZDR policies are being followed.
As such, as noted above, benchmark creators are hesitant to evaluate proprietary models on fully private test sets through model owner-hosted APIs, instead opting to create `semi-private' sets for this purpose which, while not released publicly, \emph{`suffers from a risk of leakage'} due to `exposure' to commercial APIs \cite{chollet_arc_2024}.

Additionally, since external evaluations are currently performed through voluntary agreements with model developers, evaluators have minimal bargaining power for requesting access to system components and relevant information or voicing concerns regarding their own privacy.
If model owners do not agree to the terms of engagement proposed by evaluators, they can simply walk away without repercussion. This can lead to situations where model owner privacy is stringently guarded, while evaluators are provided with bare minimum access and information, often with no solid guarantee of privacy or integrity.
However, this power imbalance could potentially flip in the case of an ill-formed regulatory requirement for external evaluation, potentially leading to a situation in which model security is significantly compromised.

\subsection{Proposed approaches do not take into account evaluator privacy}
In addition to the current practical norm being insufficient for ensuring evaluator privacy, many proposals for private, secure, or verifiable third-party evaluations do not address this shortcoming, instead focusing on the privacy of model owners.
For example, \citet{trask_how_2023} proposes a \emph{`flexible query API'} which, while allowing for the application of diverse auditing methods to private models (or training datasets) requires all audit code to be viewed and approved by the model owner.

Furthermore, while both \citet{south_verifiable_2024} and \citet{waiwitlikhit_trustless_2024} propose private and verifiable auditing processes through the application of zero-knowledge proofs (ZKPs), both assume that the model owner has full visibility into the audit being conducted.
\citet{casper_black-box_2024} propose \emph{technical}, \emph{physical}, and \emph{legal} mechanisms through which external evaluators could be provided deeper access to models, though the focus is again squarely on addressing the \emph{`risk that a developer’s models or intellectual property could be leaked'} \cite{casper_black-box_2024}.
Indeed, all mechanisms proposed can \emph{`provide white- and outside-the-box access to auditors without the system’s parameters leaving [the model developer's] servers'} -- which necessarily entails the concerns of evaluation integrity raised in the previous subsection.

Finally, while \citet{trask_secure_2024}, building on the method described in \citet{trask_how_2023}, do acknowledge the challenge of ensuring evaluator privacy, the described proof-of-concept application of secure enclaves for private evaluation allows evaluators to observe the architecture, though not the weights, of the model being evaluated.

\section{Potential solutions}
Here we outline a number of potential avenues for future work that could make progress on addressing the challenge of mutual privacy.

\subsection{Using a trusted intermediary actor}
The current, API-based approach to conducting external evaluations, as well as and the proposed solutions for ensuring model owner privacy, both assume that the model owner and the evaluator are the only two actors in the transaction.
However, one potential option for ensuring mutual privacy could be to introduce a neutral intermediary that has full visibility into the evaluation process, and can thus verify and disclose relevant information to each party.
While not reliant on technical advances, introducing such an intermediary may be politically challenging as it may require model owners to cede some amount of their current control over the evaluation process.
Furthermore, care would need to be taken to ensure that the intermediary is trustworthy and unbiased.

\subsection{Software solutions}
Progress may also be able to be made through the application of cryptographic approaches that provide provable privacy guarantees to both the model owner and external evaluator.
While there has been recent work applying techniques such as zero-knowledge proofs to private auditing \cite{south_verifiable_2024, waiwitlikhit_trustless_2024}, the focus has so far solely been on ensuring the security of the model being audited.
Future work could aim to extend this, for example, to provide cryptographic guarantees that evaluator interactions with a model through an API are not being logged by the API operator -- though the authors are not aware of any existing work aimed at achieving this.

\subsection{Hardware solutions}
The use of on-chip hardware mechanisms for private auditing is a promising area for much future work \cite{aarne_secure_2024}.
For example, and as discussed above, \citet{trask_secure_2024} outline a proof-of-concept application of secure on-chip enclaves for providing mutual privacy in evaluations.
However, this demonstration represents the early stages in the development of such methods, and thus had significant limitations. In particular, the evaluation was limited to performing benchmarking, and the specific implementation required disclosing the model architecture to the model evaluator.

% \textcolor{red}{Question for Trask: If this pilot only did benchmarking, then why was disclosing architectural details of the model necessary?}
% \textcolor{blue}{For largely uninteresting reasons. PySyft was built to be more flexible, so it naturally allows for both parties to operate using custom code, and assumes both parties want to inspect the others' code in order to approve the joint computation. It's waaay more complex to facilitate each party getting some kind of privacy/security guarnatee about the others' code without being able to read it. But this is an important topic (which is I think not covered in any paper that i know of and could be a meaty subject for this paper if you're into it)}

\section{Alternative views}
In this section we propose and discuss two alternative views in opposition to those advocated for here. Namely, we consider the views that (i) evaluators' activities should be transparent so as to enable scrutiny into conducted evaluations, and (ii) it is important for model owners to have visibility into evaluators' activities as a way of increasing the effectiveness of evaluations through collaboration.

Firstly, one could argue that information regarding external evaluations should be made visible not only to the model owner, but also to the broader public %as a way of increasing trust through facilitating scrutiny into evaluators.
in order to facilitate mechanisms for ensuring the quality and rigour of external evaluations.
% This argument echoes that given for the importance of independent, external evaluations.
While the authors find this a compelling argument, we argue that the distinction here is not binary, and that public trust in external evaluations could be furthered through means that do not impinge on the integrity of the evaluations in question.
For example, in the long-term it may be desirable to implement an `evaluator accreditation scheme' whereby an accreditation body is able to assess and approve external evaluators that meet some quality threshold.\footnote{A proto-form of such an arrangement can be seen in Article 92(2) of the AI Act which specifies that \emph{`[t]he Commission may decide to appoint independent experts to carry out evaluations on its behalf.'}}
Alternatively, a `hybrid' approach could be taken whereby the level of transparency afforded into evaluations is proportional to the necessary level of privacy. Evaluations dependent on national security-relevant information could be kept private, while more transparency could be provided for evaluations in less-sensitive domains.

Secondly, one may argue that it is important for model owners to have insight into the methods employed by external evaluators, as doing so can allow the model owner to better assist the evaluator, increasing the efficacy of the evaluation.\footnote{See, for example, the following quote regarding ARC (now known as METR) performing an audit of one of Anthropic's models: \emph{`In this case, after seeing the final audit report, we realized that we could have helped ARC be more successful in identifying concerning behavior if we had known more details about their (clever and well-designed) audit approach. [...]
% This is because getting models to perform near the limits of their capabilities is a fundamentally difficult research endeavor. Prompt engineering and fine-tuning language models are active research areas, with most expertise residing within AI companies.
With more collaboration, we could have leveraged our deep technical knowledge of our models to help ARC execute the evaluation more effectively.'} \cite{ganguli_challenges_2023}}
The authors acknowledge that in many cases model owners will have the most relevant expertise regarding how to best evaluate their model. Furthermore, this can introduce an inherent trade-off between the efficacy and independence of an external evaluation, dependent on the level of collaboration between model owner and external evaluator.
However, one could counter that many of the benefits of increased collaboration could be attained by providing external evaluators with greater access to, or additional information regarding, the model in question.
While this may carry some cost of evaluators needing to `rediscover' findings already known to the model owner, this may well be worth the additional evaluator independence afforded.

\section{Conclusion}
In this paper we have argued twofold that ensuring mutual privacy is critical for effective external evaluation of proprietary AI systems, and that both current and proposed external evaluation practices are insufficient for making such assurances.
We argue further that the issue of providing sufficient privacy guarantees to evaluators has been particularly overlooked.
Our hope is that, in highlighting this challenge, this paper will motivate future work developing technical and policy tools that can further the mutual privacy provided when conducting external evaluations, increasing their reliability, efficacy, and independence.

% Authors are kindly asked to make their submissions as accessible as possible for everyone including people with disabilities and sensory or neurological differences.
% Tips of how to achieve this and what to pay attention to will be provided on the conference website \url{http://icml.cc/}.

% \section*{Software and Data}

% If a paper is accepted, we strongly encourage the publication of software and data with the
% camera-ready version of the paper whenever appropriate. This can be
% done by including a URL in the camera-ready copy. However, \textbf{do not}
% include URLs that reveal your institution or identity in your
% submission for review. Instead, provide an anonymous URL or upload
% the material as ``Supplementary Material'' into the OpenReview reviewing
% system. Note that reviewers are not required to look at this material
% when writing their review.

\section*{Acknowledgements}

The authors would like to thank Stephen Casper, Sunishchal Dev, Matthew van der Merwe, Ishan Mishra, Andrew Trask, and participants of the Centre for the Governance of AI's \emph{Work-in-Progress} group, for insightful discussion and comments on this paper.

\section*{Impact Statement}

% Authors are \textbf{required} to include a statement of the potential 
% broader impact of their work, including its ethical aspects and future 
% societal consequences. This statement should be in an unnumbered 
% section at the end of the paper (co-located with Acknowledgements -- 
% the two may appear in either order, but both must be before References), 
% and does not count toward the paper page limit. In many cases, where 
% the ethical impacts and expected societal implications are those that 
% are well established when advancing the field of Machine Learning, 
% substantial discussion is not required, and a simple statement such 
% as the following will suffice:

This paper argues for a position that the authors feel should be discussed in the machine learning community. There are many potential societal consequences of our work, none which we feel must be specifically highlighted here.

% The above statement can be used verbatim in such cases, but we 
% encourage authors to think about whether there is content which does 
% warrant further discussion, as this statement will be apparent if the 
% paper is later flagged for ethics review.

% \newpage
\bibliography{bib}
\bibliographystyle{icml2025}

%%%%%%%%%%%%%%%%%%%%%%%%%%%%%%%%%%%%%%%%%%%%%%%%%%%%%%%%%%%%%%%%%%%%%%%%%%%%%%%
%%%%%%%%%%%%%%%%%%%%%%%%%%%%%%%%%%%%%%%%%%%%%%%%%%%%%%%%%%%%%%%%%%%%%%%%%%%%%%%
% APPENDIX
%%%%%%%%%%%%%%%%%%%%%%%%%%%%%%%%%%%%%%%%%%%%%%%%%%%%%%%%%%%%%%%%%%%%%%%%%%%%%%%
%%%%%%%%%%%%%%%%%%%%%%%%%%%%%%%%%%%%%%%%%%%%%%%%%%%%%%%%%%%%%%%%%%%%%%%%%%%%%%%
\newpage
\appendix
% \onecolumn

% You can have as much text here as you want. The main body must be at most $8$ pages long.
% For the final version, one more page can be added.
% If you want, you can use an appendix like this one.  

% The $\mathtt{\backslash onecolumn}$ command above can be kept in place if you prefer a one-column appendix, or can be removed if you prefer a two-column appendix.  Apart from this possible change, the style (font size, spacing, margins, page numbering, etc.) should be kept the same as the main body.

% \section{Related work}
% \textbf{External scrutiny, access, and transparency.}
% \cite{anderljung_towards_2023, ahmad_openais_2024,anthropic_third-party_2024,birhane_ai_2024, harrington_external_2024, casper_black-box_2024, bucknall_structured_2023, fiotto-kaufman_nnsight_2024, longpre_safe_2024, nicholas_grounding_2024, raji_outsider_2022, mokander_auditing_2023}

% \textbf{HEMs.}
% \cite{aarne_secure_2024, petrie_interim_2024, kulp_hardware-enabled_2024}

% \textbf{ZKPs.}
% \cite{south_verifiable_2024, waiwitlikhit_trustless_2024, garg_experimenting_2023, sun_zkllm_2024}

% \textbf{Model security.}
% \cite{nevo_securing_2024, openai_reimagining_2024}
%%%%%%%%%%%%%%%%%%%%%%%%%%%%%%%%%%%%%%%%%%%%%%%%%%%%%%%%%%%%%%%%%%%%%%%%%%%%%%%
%%%%%%%%%%%%%%%%%%%%%%%%%%%%%%%%%%%%%%%%%%%%%%%%%%%%%%%%%%%%%%%%%%%%%%%%%%%%%%%

\end{document}